\documentclass[review]{elsarticle}

\usepackage{hyperref}

\journal{Phisica A}

\begin{document}

\begin{frontmatter}

\title{Bridging Online and Offline Social Networks: Multiplex Analysis}

\author[manu,finki]{Sonja Filiposka}
\author[manu]{Andrej Gajduk}
\author[manu]{Tamara Dimitrova}
\author[manu,finki]{Ljupco Kocarev\corref{cor}}

\cortext[cor]{lkocarev@ucsd.edu}
\address[manu]{Macedonian Academy of sciences and Arts, Skopje, Macedonia}
\address[finki]{Faculty of Computer Science and Engineering, ``Ss Cyril and Methodius University'', Skopje, Macedonia}




\begin{abstract}
We show that three basic actor characteristics, namely normalized reciprocity, three cycles, and triplets, can be expressed using an unified framework that is based on computing the similarity index between two sets associated with the actor: the set of her/his friends and the set of those considering her/him as a friend. These metrics are extended to multiplex networks and then computed for two friendship networks generated by collecting data from two groups of undergraduate students. We found that in offline communication strong and weak ties are (almost) equally presented, while in online communication weak ties are dominant. Moreover, weak ties are much less reciprocal than strong ties. However, across different layers of the multiplex network reciprocities are preserved, while triads (measured with normalized three cycles and triplets) are not significant. 
\end{abstract}


\end{frontmatter}


\section{Introduction}

Intensity of involvement among actors in a social network and the types of actions and interactions that arise between them are a long active topic in sociological research. The analysis of group structures has started with the study of dyads and triads that was pioneered by German sociologist Georg Simmel at the end of the nineteenth century \cite{Simmel1908}. In network analysis, the importance of triads has long been emphasized in many research studies, including the highly influential work of Granovetter \cite{Granovetter1973} and, more recently, the work of Watts and Strogatz \cite{Watts1999}, where the notion of clustering (formation of many triangles in networks) is an integral part of the analysis. Thus, both social and network analysis that represent actors as networks nodes and their interactions as (un)directed links have shown that the network structural properties like reciprocity and cycles are very important when trying to explain processes like information spreading and network evolution. 

However, retaining the perspective on different connections, i.e. Granovetter's strong and weak ties, sheds additional light on the nature of connections between actors and provides deeper understanding of the triads formations and network clustering. Thus, deeper understanding of the network properties, and therefore the processes that run on top of it, can be gained if it is viewed as a collection of multiple types of links wherein each set of link types represents a separate layer of the network. This multilayered representation has led to the introduction of the concept of multiplex networks which refers to systems in which nodes are connected through more than one type of edges, and therefore, belong to multiple interacting and co-evolving networks. The importance of multiplex networks in sociology has been emphasized by many scholars. In the seminal treatment of multiple networks as the foundation of social structure, White, Boorman and Breiger  \cite{Boorman1976} and Boorman and White \cite{White1976} argued that the patterning and interweaving of different types of ties are needed to describe and characterize social structures. It has been demonstrated that multiplexity is critical to diverse phenomena, such as the mobilization of social movements \cite{Gould1991}, the consolidation of political power \cite{Padgett1993}, the emergence of trust in economic relationships \cite{Granovetter1985}, the creation of social bonds within civic networks \cite{Baldassarri2007}, and the organization of party coalitions \cite{Grossman2009}. Multiplexity has been studied to understand scientific collaboration \cite{Maggioni2013}, structural logic of intra-organizational networks \cite{Rank2010}, formation of ties featuring both an economic and a social component in inter-organizational networks \cite{Ferriani2012}, and formation of relationships among producers in the multiplex triads \cite{Shipilov2012}.

Multiplex networks have also been recently subject of particularly intense research by the network science and physics communities. Szell, Lambiotte, and Thurner \cite{Szell2010} worked on correlations and overlap between different types of links and demonstrated the tendency of individuals to play different roles in different networks. Algorithmic detection of tightly connected groups of nodes known as communities in multiplex networks was studied in \cite{Mucha2010}. A framework for growing multiplexes where a node can belong to a different networks was developed by Nicosia et al \cite{Nicosia2013}, while Kim and Goh \cite{Kim2013} studied the possibility of growth of coevolving layers that can shape the network structure and showed analytically and numerically that the coevolution can induce strong degree correlations across layers, as well as modulate degree distributions. Evolutionary game dynamics on structured populations in which individuals take part in several layers of networks of interactions simultaneously which accounts for the different kind of social ties each individual has was studied by Gomez et al \cite{Gomez2012}. 

In this paper we aim to study social relations among actors (strong and weak ties) as they appear in real life face-to face (offline) and virtual via social network sites (online) communications using the apparatus of multiplex network analysis/analytics. Real data was collected using an online survey/questionnaire given to two groups of students (two classroom based social networks). The students answers were used to map their own perception of strong vs. weak offline and online connections, thus constructing several offline and online directed friendship networks that constitute the layers of our multiplex network. The primary goal of this research is to study the interrelationship of different social structures represented as multiplex networks. For this reason, we develop normalized actor characteristics for multiplex networks, including metrics for dyads (such as reciprocity) and triads.  
We found that normalized reciprocity, three cycles, and triplets of an actor can be expressed using an unifying framework that is based on the comparison of two sets associated with the actor: the set of her/his friends - out links and the set of actors that consider her/him as a friend - in links. By extending these metrics for multiplex networks, we were able to observe the relationship of strong and weak ties in the offline and online space. Analyzing the collected data from two groups of undergraduate students we found that in offline communication strong and weak ties are (almost) equally presented, while in online communication weak ties are dominant. Moreover, weak ties are a lot less reciprocal than strong ties. However, while reciprocities are preserved across layers, the triads (measured using normalized three cycles and triplets) are not significant on the different layers of the multiplex network.

This is the outline of the paper. First, we describe the participants involved in the study and the procedure for collecting data. Next, we address the network endogenous (structural) and exogenous factors. Endogenous factors include graph characteristics such as reciprocity, three-cycles, transitive triplets, together with their normalized versions, as well as their generalizations for multiplex graphs. The section Results summarizes our findings regarding the offline and online multiplex networks generated by collecting the data described in the section Materials and Methods. We conclude the paper with the section Conclusions where we also discuss our future work.

\section{Materials and Methods}

\subsection{Participants and the procedure} 
Participants in this study were two groups of undergraduate students enrolled on two courses given at the Faculty of Computer Science and Engineering within the Cyril and Methodius University in Skopje, Macedonia. During the courses there was a dedicated lecture to explain the study and it's objectives in which the students were informed about the research experiment and only those who agreed to participate were part of the experiment. Thus, the two groups represent two social networks of students that attend the same class during one semester of their studies. The total number of students in the first group, Group 1, is 171, out of which 153 voluntarily participated in the study. The gender distribution is almost equal; there were 86 female and 85 male students. The total number of students in the second group, Group 2, is 150 (118 male, 32 female), while only 80 participated in the At the end we summarize our findings regarding the offline and online multiplex networks generated from the collected datastudy. The students age for both groups ranges from 19 years to 22 years. Additional data was also collected for each student via the university online electronic course enrollment system (such as: gender, study program, GPA, accumulated credits). 

The online survey was developed by the research team. Each student (that participated in the study) was asked to select from the presented list of classmates those with whom she/he was engaged in face-to-face (offline) communication and Facebook (online) communication. Each chosen contact was described as weak or strong according to the student's own perception by the means of pre-defined categories for frequency of communication. In the case of face-to-face communication a contact (tie) is considered to be strong if the frequency of communication is higher than 5 interactions per month, while it is considered as weak contact (tie) if the number of interactions is larger than once in 3 months, but less than 5 interactions per month. When considering the online communication, strong contacts (ties) are those with whom the student interacts more than five times per week, while a weak contact (tie) is the one with whom there was Facebook activity more than once a month, but less than five times per week. Facebook was chosen as the online social network representative since it's use is extremely wide spread among the population.

\subsection{Friendship graphs as multiplex networks}

Graphs provide a powerful primitive for modeling data in social science. Nodes usually represent real world objects and edges indicate relationships between objects. In sociology, nodes may have attributes associated with them and graphs may contain many different types of relationships. The node attributes are used to describe the features of the objects that the nodes represent. For example, a node representing a student may have attributes that represent the student's gender and department. Different types of edges in a graph correspond to different types of relationships between nodes, such as friends and classmates relationships. 

Here we study friendship relations among $n$ actors: the existence of a tie $i \to j$ will be described as $i$ calling $j$ a friend. The ties are represented as binary variables, denoted by $x_{ij}$. A tie from actor $i$ to actor $j$, $i \to j$, is either present or absent ($x_{ij}$ then having values 1 and 0, respectively). The tie variables constitute the network, represented by its $n \times n$ adjacency matrix $X = [x_{ij}]$ (self-ties are excluded). The graph is directed, where each tie $i \to j$ has a sender $i$, who will also be referred to as ego, and a receiver $j$, referred to as alter, as it is common in social network analysis. 

We model different types of relationships among actors using the concept of multiplex graphs. In sociology, multiplex graphs (networks) refer to the case when nodes (actors) are connected through more than one type of (socially relevant) ties. In mathematics, such graphs are also called multi-graphs  (a multi-graph is a graph that is allowed to have multiple edges, that is, edges that have the same end nodes). We give a definition of multiplex graphs adapted for the study reported in this paper. Let $V $ denote the set of nodes; nodes are connected via $L$ different type of connections (ties). Each type of connection (together with the set of nodes) forms a (directed) graph with $n$ vertices. We denote with $G^\alpha = (V, E^\alpha)$ the graph which represents the connection type $\alpha$, where $E^\alpha$ denotes the set of $\alpha$-type ties, $\alpha = 1, \ldots, L$. Let $ [x_{ij}(G^\alpha)]$ be $n \times n$ adjacency matrix of the graph $G^\alpha$. A multiplex graph $\cal{G}$ is then defined as a collection of all graphs $G^\alpha$ and all edge-aggregated graphs of the form $(V, \cup_{\alpha =\alpha_1}^{\alpha_m} E^\alpha)$, where $\alpha_1, \alpha_m \in \{1, \ldots, L\} $. We assume that the case $\alpha_1=\alpha_m$ is not excluded and thus write
$$
{\cal{G}}  = \{ G^{\alpha_1 \alpha_2 \ldots \alpha_m} = (V, E^{\alpha_1} \cup E^{\alpha_2} \cup \ldots \cup E^{\alpha_m}): \alpha_1, \ldots \alpha_m \in \{1, \ldots, L\} \}
$$
for the multiplex graph $\cal{G}$. In the study discussed here, different types of edges correspond to the four types of friendship relations: online weak or strong connections and offline weak or strong connections.

\subsection{Jaccard similarity coefficient} 

In order to introduce an unifying framework for discussing graph characteristics, we use the Jaccard index, also known as Jaccard similarity coefficient. It measures the similarity between finite sets and is defined as the size of the intersection divided by the size of the union of the sets:
$$
J(A,B) = \frac{|A \cap B|}{|A \cup B|} = \frac{|A \cap B|}{ |A| + |B| - |A \cap B| }, 
$$
where $\left\vert{S}\right\vert$ denotes the cardinality of the set $S$. If $A$ and $B$ are both empty, $J(A,B)$ is defined as 1. Note that
$$
0\le J(A,B)\le 1. 
$$  
The Jaccard distance, which measures dissimilarity between sets and is a metric on the collection of all finite sets, is complementary to the Jaccard coefficient and is obtained by subtracting the Jaccard coefficient from 1. 
Using Jaccard index to introduce various actor-based characteristics is based on the following observation. Let $A_i = \{a: i \to a  \}$ and $B_j = \{b: j \to b   \}$. Then   
\begin{eqnarray*}
|A_i \cap B_j| & = &  \sum_h x_{ih} x_{jh}    \\
|A_i \cup B_j| &= & \sum_h x_{ih} + \sum_h x_{jh}  -  \sum_h x_{ih} x_{jh},  
\end{eqnarray*}
where $|A_i|  = \sum_h x_{ih} $ and $ |B_j| = \sum_h x_{jh}$.

\subsection{Network (endogenous) characteristics}

We consider a number of social network characteristics; those depending only on the network are called structural or endogenous characteristics, while characteristics depending on externally given attributes are called covariate or exogenous characteristics.

\subsubsection{Simple-graph characteristics} 

Let $G=(V,E)$ be a directed graph and let $i$ be an arbitrary node (actor) in the graph. We denote $S^{in}_i(k)$ and $S^{out}_i(k)$ for the $k$-in-neighborhood of $i$ and $k$-out-neighborhood of $i$, respectively, i.e., for the set of vertices from where vertex $i$ can be reached and for the set of vertices that can be reached from $i$ in $k$ steps (using $k$ directed edges). More formally, these sets are defined as 
\begin{eqnarray}
S^{in}_i(k) & = & \{j:  \mbox{ there is a path of length } k \mbox{ from } j \mbox{ to } i      \} \label{set-in} \\
S^{out}_i(k) & = & \{j:      \mbox{ there is a path of length } k \mbox{ from } i \mbox{ to } j     \}  \label{set-out}  
\end{eqnarray}
The general case of arbitrary $k$ will be discussed elsewhere, here we consider only to the case when $k=1$. For $k=1$ we use the shorter notations $S^{in}_i(1) = S^{in}_i$ and $S^{out}_i(1) = S^{out}_i$. In other words, we consider the 1-hop-neighbor sets of node $i$ defined as:   
\begin{eqnarray}
S^{out}_i &=& \{j:  i \to j  \}, \label{eq-s1}  \\
S^{in}_i  &=& \{j: j \to i  \}, \label{eq-s2} 
\end{eqnarray}
If the existence of a tie $i \to j$ is being interpreted as $j$ is a friend of $i$, then $S_i^{out}$ is the set of $i$'s friends, while $S_i^{in}$ is the set of nodes that consider $i$ as their friend. For the cardinalities of these sets, we write $d^{out}_i  = \sum_j x_{ij} = |S^{out}_i|$ and $d^{in}_i    = \sum_j x_{ji} = |S^{in}_i|$. Therefore, $d_i^{out}$ is the number of $i$'s friends and $d_i^{in}$ is the number of nodes that consider $i$ as their friend. These two metrics are known as activity, i.e. tendency to establish friendships, and popularity, i.e. ability to gain friends, in social science. In this paper, we analyse several graph characteristics that can all be expressed using the Jaccard similarity index for a given pair of node sets. Out of the many different possible characteristics, here the focus is set on those that we consider most relevant for this study: reciprocity, three-cycles, and triplets.

\textit{Reciprocity} --
One of the most basic properties of social networks is reciprocity, represented by the number of reciprocated ties of actor $i$ and defined as $\sum_j x_{ij} x_{ji}$, which can be rewritten as 
$$
Rec_i = |S^{out}_i \cap S^{in}_i| = \sum_j x_{ij} x_{ji}
$$
Here we consider 
normalized reciprocity defied as:  
\begin{equation} \label{eq-r1}
r_i = J\left(S_i^{out},S_i^{in}\right) =  \frac{\sum_j x_{ij}x_{ji}}{  d_i^{out} + d_i^{in} - \sum_j x_{ij}x_{ji} }. 
\end{equation}
The normalized reciprocity enables one to decide on the quantity of reciprocated ties when compared to the total number of ties of both actors. This normalization provides a perspective and places the phenomenon of reciprocity into context by comparing it to the total number of possibilities for reciprocated ties under the given circumstances.
\textit{Three-cycles} -- Next to reciprocity, an essential feature in most social networks is transitivity, or transitive closure which is represented by two metrics: three-cycles and triplets. A cycle of length 3 in a given graph is defined as a sub-graph that consists of a sequence of directed edges $i \to j \to h \to i$ which connect a sequence of vertices $i, j, h$, all distinct from one another. 
We discuss the first metric, three-cycles, defined as the number of three-cycles an actor $i$ is involved in, that is 
$$
cyc_i = \sum_{j,h} x_{ij} x_{jh} x_{hi}
$$
We introduce a normalized characteristic for the number of three-cycles an actor $i$ is involved in as follows: 
\begin{eqnarray*}
tc_i & = &  \frac{1}{d_i^{in}} \sum_{h} x_{hi} J(S_i^{out},S_h^{in}) \\
& = &  \frac{1}{d_i^{in}} \sum_{h} \frac{ x_{hi} \sum_j x_{ij} x_{jh}  }{d_i^{out} +d_h^{in} - \sum_j x_{ij} x_{jh} }
\end{eqnarray*} 
$| S_i^{out} \cap  S_h^{in}| $ is the number of the common neighbors for both $i$ and $h$ or, the number of $i$'s friends that consider $h$ to be a friend as well.  
The Jaccard index between these two sets $S_i^{out}$ and $S_h^{in}$ reflects how similar these sets are in terms of common versus non-common friends. Moreover,  
$$
\sum_h x_{hi} | S_i^{out} \cap  S_h^{in}| = \sum_h x_{hi} \sum_j x_{ij} x_{jh} 
$$ 
is the number of three-cycles that actor $i$ is involved in.  Therefore, by summarizing the Jaccard indexes $J(S_i^{out},S_h^{in})$ for all $h $ 
one can obtain a measure (metric) $tc_i$ that represents the status of common versus non-common friends of $i$ in the graph when considered in terms of transitive three-cycles.

\textit{Transitive triplets} --  
A transitive triplet of length 3 in a given graph is a sub-graph that consists of a sequence of directed edges $i \to j \to h$ and $ i \to h$ which connect a sequence of vertices $i, j, h$ that are all distinct from one another. Recall, the metric triplets is defined as the number of triplets an actor $i$ is involved in, that is 
$$
plt_i = \sum_{j.h} x_{ij} x_{jh} x_{ih}.
$$
In a similar fashion as for the three-cycles, one can define normalized metric for the number of triplets in the graph: 
\begin{eqnarray*}
tp_i & = & \frac{1}{d_i^{out}} \sum_{j} x_{ij} J(S_i^{out},S_j^{out}) \\
& = &  \frac{1}{d_i^{out}} \sum_{j}  \frac{ x_{ij} \sum_h x_{ih} x_{jh}  }{d_i^{out} + d_j^{out} - \sum_h x_{ih} x_{jh}   } 
\end{eqnarray*} 
Note that the normalized reciprocity, normalized transitive triplets, and normalized three-cycles are related to reciprocity, transitive triplets, and three-cycles, respectively - quantities that are commonly used in social science and graph theory. The fact, shown here, that they can be expressed with Jaccard similarity index not only brings novel understanding of these quantities, but also suggests how they can be extended for multiplex networks, which will be done in the next section. However, take into consideration that while using the Jaccard similarity index to express the network characteristics, we define it as 0 when both considered sets are empty in order to reflect the social aspects of the interpretation, i.e. two nodes that have no friends at all are not reciprocated and are not part of any triplets or cycles.

\subsubsection{Multiplex-graph characteristics}

When considering a multiplex graph wherein $\alpha$ and $\beta$ ($\alpha = \beta $ is not excluded) represent two types of social relationships, the normalized reciprocity, normalized three-cycles, and normalized transitive triplets  can be generalized as:
\begin{eqnarray*} 
r_i(G^\alpha, G^\beta) &=& J\left(S_i^{out} (G^\alpha) ,S_i^{in} (G^\beta) \right) \\
tc_i(G^\alpha, G^\beta) &=&  \frac{1}{d_i^{in}} \sum_{h} x_{hi} J(S_i^{out}(G^\alpha),S_h^{in}(G^\beta)) \\
tp_i (G^\alpha, G^\beta) &=&  \frac{1}{d_i^{out}} \sum_{j} x_{ij} J(S_i^{out}(G^\alpha),S_j^{out}(G^\beta)).  
\end{eqnarray*}
By using these metrics we can analyze how different types of links interact and form mixed dyad and triad formations. Recalling that in the general case a multiplex graph can consist of several different layers, the multiplex-graph extension of the reciprocity, cycles and triplets metrics enables us to study what type of links are stronger than others, as well as how the network characteristics change when considering only a subset of all existing interrelations among its nodes. 

We also consider the following two characteristics called overlapping indexes and defined as:    
\begin{eqnarray*}
oi_i^{out} (G^\alpha, G^\beta) &= & J(S_i^{out}(G^\alpha), S_i^{out}(G^\beta))  \\
oi_i^{in} (G^\alpha, G^\beta) & = & J(S_i^{in}(G^\alpha), S_i^{in}(G^\beta)) 
\end{eqnarray*}
These two metrics are introduced in order to create a contrasting view compared to the normalized multiplex graph characteristics as they are defined above. By analyzing the Jaccard similarity of sets of links for a given node that belong to different layers of the multiplex graph, we are able to infer the consistency of the relationship intensity for different types of links. In other words, does the node have the tendency to have different types of relationship with the same set of friends in different environments (for example, online and offline environment in our study case). We argue that the out overlapping index shows the node activity in creating and maintaining different types of links, while the in overlapping index represents the node popularity, i.e. ability to gain different types of relationships.

\subsection{Exogenous characteristics} 

Nodes in a graph may have a set of associated attributes, also called exogenous actor covariates. In order to emphasize the completeness of the approach, in this section we show that our unified framework based on the Jaccard similarity index can also be used for the analysis of the exogenous characteristics of the multiplex graph. The results for the exogenous characteristics will be presented elsewhere. 

Let $A_i$ be the set of attributes associated to the node $i$. There are two basic characteristics for the actor $i$: the out-attributes characteristic, measuring whether actors with higher similarity-index values tend to nominate more friends and hence have a higher out-degree; the in-attributes characteristic, measuring whether actors with higher similarity-index values will tend to be nominated by many others and hence have higher in-degrees. These two characteristics are defined as 
\begin{eqnarray}
att^{out}_i &=& \frac{1}{d_i^{out}} \sum_j x_{ij} J(A_i, A_j) \label{eq-eq} \\
att^{in}_i &=&\frac{1}{d_i^{in}} \sum_j x_{ji} J(A_i, A_j) \label{eq-al}, 
\end{eqnarray}
which will be compared to the average un-networked similarity value computed as 
$$
att = \frac{2}{n(n-1)} \sum_{i,j; i<j} J(A_i, A_j).  
$$
The characteristic $att^{out}_i$ measures similarity of attributes of the pair of end-nodes $i$ and $j$ for all  the friends of $i$, that is, for all nodes in the set $S_i^{out}$. The characteristic $att^{in}_i$ measures similarity of attributes of the pair of end-nodes $i$ and $j$ for those $j$ that consider $i$ as a friend, that is, for all nodes in the set $S_i^{in}$.  The quantities (\ref{eq-eq}) and (\ref{eq-al}) can be extended for multiplex networks as    
\begin{eqnarray*}
att^{out}_i(G^\alpha) &=& \frac{1}{d_i^{out}} \sum_j x_{ij} (G^\alpha) J(A_i, A_j) \\
att^{in}_i (G^\alpha) &=&\frac{1}{d_i^{in}} \sum_j x_{ji} (G^\alpha) J(A_i, A_j)  
\end{eqnarray*}
With this extension once can also analyze the similarity of the attributes for different pairs of nodes that are connected across different layers of the multiplex network. Therefore, these metrics provide insight into how different types of links influence the attribute based node similarity.

\section{Results}

\begin{figure}[h] \label{fig-1}
\begin{center}
	\includegraphics[width=\textwidth]{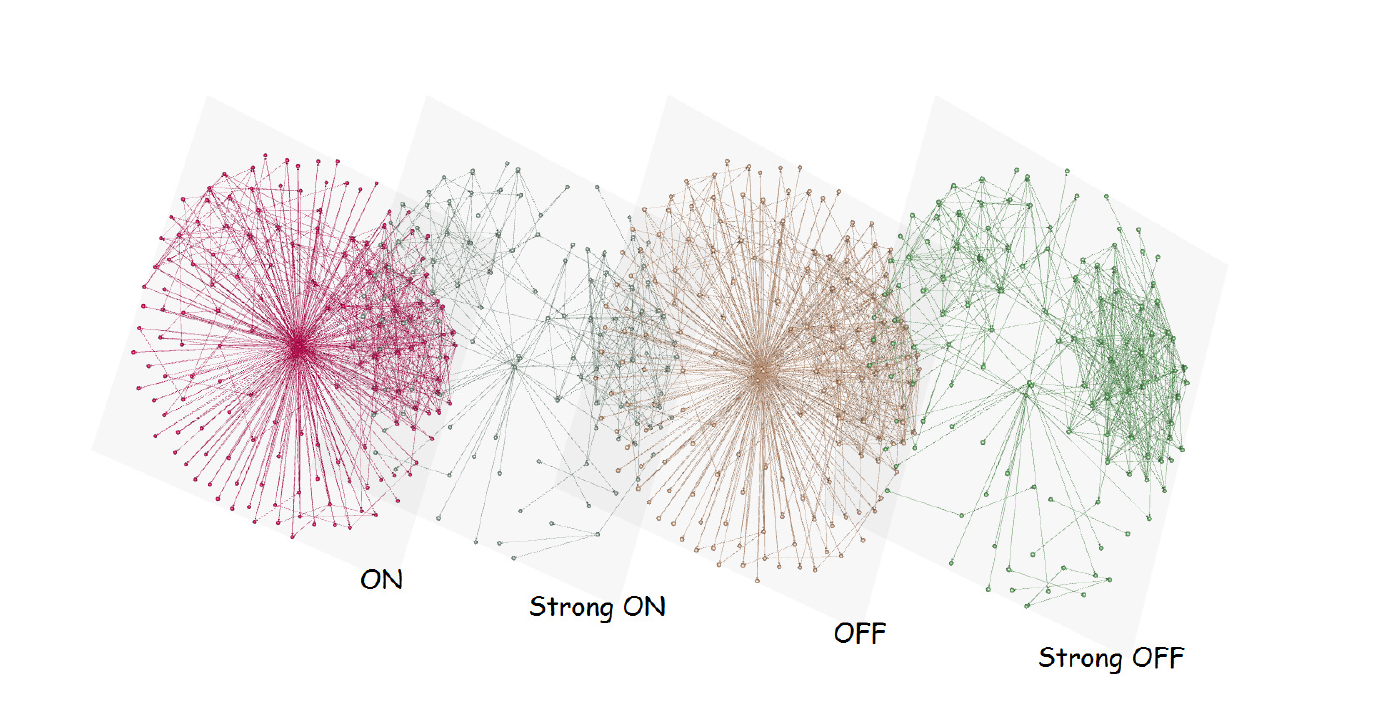} 
	\caption{{\bf Visualization of a part of the multiplex graph constructed according to the data from Group 2 with 2 basic and 2 aggregated graphs. Note that all nodes (if present) are in the same positions across the layers.}}
\end{center}
\end{figure}

Let $V$ be the set of all students enrolled on a given course (group). For each group of students, there are four distinct directed graphs that have been generated based on the answers collected from the online survey: offline/online social network with strong ties and offline/online social network with weak ties, defined as: 
\begin{eqnarray}
G^{of}_s & = & (V, E^{of}_{s}),  \hspace{0.5cm} E^{of}_{s} = \{ i \to j \mbox{ is an offline strong tie} \} \label{sOFF} \\
G^{of}_w & = & (V, E^{of}_{w}), \hspace{0.5cm} E^{of}_{w} = \{ i \to j \mbox{ is an offline weak tie} \} \label{wOFF} \\
G^{on}_s & = & (V, E^{on}_{s}), \hspace{0.5cm} E^{on}_{s} = \{ i \to j \mbox{ is an online strong tie} \} \label{sON} \\
G^{on}_w & = & (V, E^{on}_{w}), \hspace{0.5cm} E^{on}_{w} = \{ i \to j \mbox{ is an online weak tie} \} \label{wON} 
\end{eqnarray}
From these four (basic) graphs, five more aggregated graphs are constructed: 
\begin{eqnarray}
G^{of} & = & (V, E^{of}), \hspace{0.5cm} E^{of} = E^{of}_{s} \cup E^{of}_{w} \label{OFF} \\
G^{on} & = & (V, E^{on}), \hspace{0.5cm}   E^{on} = E^{on}_{s} \cup E^{on}_{w} \label{ON} \\
G_s & = & (V, E_s), \hspace{0.5cm}   E_s = E^{on}_{s} \cup E^{of}_{s} \label{s} \\
G_w & = & (V, E_w), \hspace{0.5cm}   E_w = E^{on}_{w} \cup E^{of}_{w} \label{w} \\
G & = & (V,E)  \hspace{0.5cm} E = E^{of} \cup E^{on} = E_s \cup E_w \label{all}
\end{eqnarray}

Two multiplex networks are studied for two groups. Each network consists of nine layers or nine sets of ties: four basic - strong OFF, weak OFF, strong ON, and weak ON and five aggregated sets - strong, weak, ON, OFF, and all, as described with the given equations. In Fig. 1 a partial visual representation of the multiplex network for Group 2 is presented aiming to conceptualize the different types of links in each layer together with the possible ways for aggregation.

\begin{table}[!ht] \label{tab-1}
\caption{{\bf Basic graph characteristics for all layers in the multiplex graph that represents the social interrelations between the students from Group 1}}
\begin{tabular}{|l|l|l|l|l|l|l|l|l|} 
\hline
 Group 1 & { $|V|$}&{ $|E|$}&{ TotDegree}&{ Assor}&{ $|V_1|$}&{ $|E_1|$ }&{ Path }&{ Diam}\\ \hline
{ strong OFF}&{ 153}&{ 675}&{ 4.412}&{ 0.431}&{ 122}&{ 630}&{ 4.632}&{ 14}\\ \hline
{ weak OFF}&{ 153}&{ 756}&{ 4.941}&{ 0.018}&{ 135}&{ 693}&{ 3.856}&{ 9}\\ \hline
{ OFF}&{ 153}&{ 1420}&{ 9.281}&{ 0.238}&{ 150}&{ 1402}&{ 2.994}&{ 7}\\ \hline
{ strong ON}&{ 153}&{ 428}&{ 2.797}&{ 0.235}&{ 87}&{ 319}&{ 5.196}&{ 14}\\ \hline
{ weak ON}&{ 153}&{ 783}&{ 5.118}&{ 0.154}&{ 138}&{ 742}&{ 3.782}&{ 9}\\ \hline
{ ON}&{ 153}&{ 1201}&{ 7.850}&{ 0.253}&{ 150}&{ 1192}&{ 3.251}&{ 7}\\ \hline
{ strong}&{ 153}&{ 730}&{ 4.771}&{ 0.392}&{ 128}&{ 689}&{ 4.339}&{ 11}\\ \hline
{ weak}&{ 153}&{ 1046}&{ 6.837}&{ 0.087}&{ 145}&{ 1016}&{ 3.212}&{ 7}\\ \hline
{ all}&{ 153}&{ 1487}&{ 9.719}&{ 0.195}&{ 153}&{ 1487}&{ 2.937}&{ 7}\\ \hline
\end{tabular}
\end{table}

\begin{table}[!ht] \label{tab-2} 
\caption{{\bf Basic graph characteristics for all layers in the multiplex graph that represents the social interrelations between the students from Group 2}}
\begin{tabular}{|l|l|l|l|l|l|l|l|l|}
\hline
{  Group 2 }&{ $|V|$}&{ $|E|$}&{ TotDegree}&{ Assor}&{ $|V_1|$ }&{ $|E_1|$ }&{ Path }&{ Diam}\\ \hline
{ strong OFF}&{ 80}&{ 412}&{ 5.150}&{ 0.184}&{ 64}&{ 390}&{ 3.337}&{ 9}\\ \hline
{ weak OFF}&{ 80}&{ 521}&{ 6.513}&{ 0.297}&{ 57}&{ 420}&{ 2.697}&{ 5}\\ \hline
{ OFF}&{ 80}&{ 930}&{ 11.625}&{ 0.360}&{ 71}&{ 892}&{ 2.282}&{ 5}\\ \hline
{ strong ON}&{ 80}&{ 226}&{ 2.825}&{ 0.199}&{ 49}&{ 188}&{ 4.468}&{ 12}\\ \hline
{ weak ON}&{ 80}&{ 465}&{ 5.812}&{ 0.352}&{ 62}&{ 414}&{ 2.922}&{ 7}\\ \hline
{ ON}&{ 80}&{ 690}&{ 8.625}&{ 0.356}&{ 68}&{ 651}&{ 2.544}&{ 5}\\ \hline
{ strong}&{ 80}&{ 455}&{ 5.688}&{ 0.213}&{ 65}&{ 430}&{ 3.157}&{ 9}\\ \hline
{ weak}&{ 80}&{ 728}&{ 9.100}&{ 0.352}&{ 67}&{ 675}&{ 2.463}&{ 5}\\ \hline
{ all}&{ 80}&{ 1013}&{ 12.662}&{ 0.384}&{ 73}&{ 985}&{ 2.243}&{ 5}\\ \hline
\end{tabular}
\end{table}

Multiplex networks are particularly significant when they overlap and interact to create phenomena or processes that cannot be explained by a single network alone. Tables 1 and 2 summarize the basic graph characteristics for all graphs generated in the study (total of 9 graphs per group): number of actors $|V|$, number of ties $|E|$, average degree and assortativity.  Recall, a directed graph is strongly connected if there is a directed path from each vertex to every other vertex. The strongly connected components (SCC) of a directed graph are its maximal strongly connected sub-graphs. The number of actors $|V_1|$ and ties $|E_1|$ for the largest SCCs for each of these graphs are also presented in the tables, including the average path length and the diameter. 

Since the number of participants in the two groups (153 versus 80) is different, it is significant to confirm that the same conclusions (especially concerning the ratios) hold for both groups. Namely, the number of strong ties is smaller than the number of weak ties: 689 versus 1016 (40\% strong ties and 60\% weak ties) for the first group and 430 versus 685 (41\% strong ties and 59\% weak ties) for the second group. However, if we take a closer look at the strong ties graphs, one can notice that there are more strong offline ties than strong online ties. This leads us to the conclusion that the students have closer friendship relations in the offline real rather than the online virtual environment. The number of ON ties is smaller than the number of OFF ties: 1192 versus 1402 (46\% ON ties and 54\% OFF ties) for Group 1 and 651 versus 892 (42\% ON ties and 58\% OFF ties) for Group 2. Moreover, the number of strong OFF ties is almost the same as the number of weak OFF ties for both groups: 630 and 693 (48\% and 52\%) for the first group and 390 and 420 (48\% and 52\%) for the second group. This suggests that weak communications are (almost) equally presented in the online and real life communication. However, the number of strong ON ties is almost half the number of weak ON ties: 319 and 742 (30\% and 70\%) for the first group and 188 and 414 (31\% and 69\%) for the second group. This could be interpreted as the fact that students within one group interact with all colleagues no matter whether they consider them close or distant which could be due to the necessities of working together on different projects, homework or labs for example. However, looking at the number of online strong and online weak ties, we can infer that the students use the virtual world to spread weak and more common friendships most probably aiming to expand their circle of acquaintances.  

Comparing the average total degrees between basic graphs and aggregated graphs in the multiplex network it is fairly straightforward to conclude that both groups exhibit similar patterns for the average degrees: strong OFF and weak OFF graphs have average degrees 4.4 and 4.9 for the first group and 5.1 and 6.5 for the second group, respectively. Strong ON and weak ON graphs have average degrees 2.8 and 5.1 for the first group and 2.8 and 5.8 for the second group, respectively. At the aggregated level, the average degrees for strong and weak graphs are 4.8 and 6.8 for the first group and 5.7 and 9.1 for the second group, while for the OFF and ON graphs these numbers are 9.3 and 7.8 for the first group and 11.6 and 8.6 for the second group. Note that the average degrees for all graphs in the multiplex network of the second group are greater than the average degrees for the corresponding graphs of the first group.  One possible explanation is that the students which are part of smaller group are more friendly and associative between them. This is especially the case in this scenario since the the students belonging to Group 2 have a more diverse background (i.e. type of study program, year of study and alike) compared to Group 1. However, we do not have more data to confirm (or disconfirm) this conclusion (hypothesis).

Tables 3 and 4 provide summaries of the basic metrics: average values for the reciprocities, three cycles, and triplets for all (single) graphs in the multiplex networks that represent group 1 and 2, respectively. All weak graphs (weak OFF, weak ON, and weak) have smaller values for reciprocity, three cycles, and triplets for both groups indicating that weak ties are less socially significant.  Both transitive triplets and three-cycles represent closed structures, however, triplets indicate hierarchical ordering in contrast to three-cycles which are against of such ordering. For all graphs studied here, the average values of transitive triplets is slightly larger than the average values of three-cycles showing that the elements of hierarchical ordering are present in these social networks.  Also, the results given in both tables 3 and 4 indicate that there is a correlation between dyads (reciprocity) and triads (transitive triplet and three-cycle) such that larger (smaller) values of the former imply larger (smaller) values of the latter.

\begin{table}[!ht]
\caption{{\bf Basic endogenous characteristics for all layers in the multiplex network that represents social interactions in Group 1}}
\begin{tabular}{|l|l|l|l|}
\hline
{\bf Group 1 }& Average $r_i$  & Average $tc_i$ & Average  $tp_i$ \\ \hline
{ strong OFF}&{ 0.445}&{ 0.153}&{ 0.168}\\ \hline
{ weak OFF}&{ 0.154}&{ 0.029}&{ 0.039}\\ \hline
{ OFF}&{ 0.467}&{ 0.147}&{ 0.190}\\ \hline
{ strong ON}&{ 0.465}&{ 0.141}&{ 0.136}\\ \hline
{ weak ON}&{ 0.204}&{ 0.045}&{ 0.049}\\ \hline
{ ON}&{ 0.442}&{ 0.133}&{ 0.173}\\ \hline
{ strong}&{ 0.482}&{ 0.159}&{ 0.177}\\ \hline
{ weak}&{ 0.278}&{ 0.061}&{ 0.074}\\ \hline
{ all}&{ 0.492}&{ 0.155}&{ 0.200}\\ \hline
\end{tabular}
\end{table}

\begin{table}[!ht]
\caption{{\bf Basic endogenous characteristics for all layers in the multiplex network that represents social interactions in Group 2}}
\begin{tabular}{|l|l|l|l|}
\hline
{Group 2 }& Average $r_i$  & Average $tc_i$ & Average  $tp_i$ \\ \hline
{ strong OFF}&{ 0.425}&{ 0.156}&{ 0.161}\\ \hline
{ weak OFF}&{ 0.074}&{ 0.041}&{ 0.049}\\ \hline
{ OFF}&{ 0.335}&{ 0.139}&{ 0.200}\\ \hline
{ strong ON}&{ 0.448}&{ 0.146}&{ 0.149}\\ \hline
{ weak ON}&{ 0.099}&{ 0.049}&{ 0.064}\\ \hline
{ ON}&{ 0.342}&{ 0.115}&{ 0.172}\\ \hline
{ strong}&{ 0.509}&{ 0.150}&{ 0.171}\\ \hline
{ weak}&{ 0.148}&{ 0.081}&{ 0.102}\\ \hline
{ all}&{ 0.360}&{ 0.153}&{ 0.216}\\ \hline
\end{tabular}
\end{table}

Tables 5 and 6 show the average values of 5 different additional characteristics that are focusing on the interrelationship of different parts of the multiplex graph. Here, in addition to reciprocity, three-cycle, and triplet, we also have the two overlapping indexes, all as they are defined in the subsection Multiplex graph characteristics. The given results indicate that reciprocity is preserved across different layers of the multiplex network; in particular the ties in strong OFF are reciprocal with ties in strong ON and vice versa (the normalized average values are $r_i(G_s^{OFF}, G_s^{ON}) = 0.417$ and  $r _i(G_s^{ON}, G_s^{OFF}) = 0.406$, respectively. Similar values are also obtained for the pairs (OFF, ON) and (ON, OFF). However, for the considered social multiplex network, the triads (measured with normalized three cycles and triplets) are not significant. On the other hand, the values for the overlapping indexes show that activity and popularity patterns among some layers of the multiplex network are significant. For instance, the number of out-degree and in-degree friends in the strong OFF layer coincides with the out-degree and in-degree friends in the strong ON layer. Or, out-degree and in-degree friends in the weak OFF layer are also out-degree and in-degree friends in the weak ON layer. On the other hand, the overlapping indexes for both out-degree (activity) and in-degree (popularity) are small for the following combinations of two graphs (strong ON, weak ON) and (strong OFF, weak, OFF) for both Groups 1 and 2.

\begin{table}[!ht]
\caption{{\bf Multiplex-graph characteristics for different pairs of layers reflecting the combined types of ties for Group 1}}
\begin{tabular}{|l|l|l|l|l|l|}
\hline
{\bf Group 1}&{\bf Reciprocity}&{\bf tc2}&{\bf tp2} & oi1 & oi2 \\ \hline
{ strong OFF, strong ON}&{ 0.417}&{ 0.136}&{ 0.148}  &{ 0.568}&{ 0.524}  \\ \hline
{ weak OFF, weak ON}&{ 0.160}&{ 0.032}&{ 0.037}    &{ 0.427}&{ 0.512}     \\ \hline
{ strong OFF, weak OFF}&{ 0.091}&{ 0.061}&{ 0.065}   &{ 0.025}&{ 0.005}  \\ \hline
{ strong ON, weak ON}&{ 0.081}&{ 0.052}&{ 0.059}   &{ 0.028}&{ 0.006}    \\ \hline
{ strong, weak}&{ 0.175}&{ 0.093}&{ 0.102}     &{ 0.195}&{ 0.180}         \\ \hline
{ OFF, ON}&{ 0.434}&{ 0.141}&{ 0.182}      &{ 0.735}&{ 0.775}            \\ \hline
{ strong ON, strong OFF}&{ 0.406}&{ 0.145}&{ 0.136}  &{ 0.568}&{ 0.524}   \\ \hline
{ weak ON, weak OFF}&{ 0.165}&{ 0.038}&{ 0.042}    &{ 0.427}&{ 0.512}     \\ \hline
{ weak OFF, strong OFF}&{ 0.116}&{ 0.040}&{ 0.061}  &{ 0.025}&{ 0.005}  \\ \hline
{ weak ON, strong ON}&{ 0.098}&{ 0.046}&{ 0.064}   &{ 0.028}&{ 0.006}    \\ \hline
{ weak, strong}&{ 0.207}&{ 0.075}&{ 0.098}    &{ 0.195}&{ 0.180}        \\ \hline
{ ON, OFF}&{ 0.433}&{ 0.139}&{ 0.179}   &{ 0.735}&{ 0.775}                \\ \hline
\end{tabular}
\end{table}


\begin{table}[!ht]
\caption{{\bf Multiplex-graph characteristics for different pairs of layers reflecting the combined types of ties for Group 2}}
\begin{tabular}{|l|l|l|l|l|l|}
\hline
{\bf Group 2}&{\bf Reciprocity}&{\bf tc2}&{\bf tp2}   & oi1 & oi2      \\ \hline
{ strong OFF, strong ON}&{ 0.411}&{ 0.127}&{ 0.138}  &{ 0.517}&{ 0.470}   \\ \hline
{ weak OFF, weak ON}&{ 0.072}&{ 0.038}&{ 0.057}   &{ 0.400}&{ 0.401}      \\ \hline
{ strong OFF, weak OFF}&{ 0.071}&{ 0.054}&{ 0.056}  &{ 0.093}&{ 0.003}    \\ \hline
{ strong ON, weak ON}&{ 0.062}&{ 0.042}&{ 0.047}   &{ 0.102}&{ 0.001}     \\ \hline
{ strong, weak}&{ 0.132}&{ 0.086}&{ 0.095}     &{ 0.269}&{ 0.141}         \\ \hline
{ OFF, ON}&{ 0.334}&{ 0.117}&{ 0.181}      &{ 0.708}&{ 0.604}               \\ \hline
{ strong ON, strong OFF}&{ 0.362}&{ 0.130}&{ 0.138}   &{ 0.517}&{ 0.470}  \\ \hline
{ weak ON, weak OFF}&{ 0.067}&{ 0.040}&{ 0.047}   &{ 0.400}&{ 0.401}      \\ \hline
{ weak OFF, strong OFF}&{ 0.110}&{ 0.041}&{ 0.070}  &{ 0.093}&{ 0.003}     \\ \hline
{ weak ON, strong ON}&{ 0.135}&{ 0.039}&{ 0.065}  &{ 0.102}&{ 0.001}       \\ \hline
{ weak, strong}&{ 0.209}&{ 0.073}&{ 0.103}    &{ 0.269}&{ 0.141}           \\ \hline
{ ON, OFF}&{ 0.306}&{ 0.118}&{ 0.180}  &   { 0.708}&{ 0.604}              \\ \hline
\end{tabular}
\end{table}

In social science structural equivalence is defined as ``two nodes are considered structurally equivalent if they share many of the same network neighbors.'' A possible operationalization of this definition could be done as follows: two nodes $i$ and $j$ are structurally equivalent if $r_i = r_j$, $tc_i = tc_j$ and $tp_i = tp_j$. We found that in each network some of the actors are structurally equivalent. For example in the strong ON graph for Group 1, the actors with id 2, 24, and 127 are structurally equivalent having $r_i=1$,  $tc_i = 0.55$, and $tp_i = 0.532$. These actors have out-degree and in-degree $d_i^{out} = 4$ and $d_i^{in} = 4$, respectively. Table 7 shows characteristics for those actors in the strong OFF graph that have $d_i^{out} = 4$ and $d_i^{in} = 4$. Again some of those actors are structurally equivalent (2 and 127). These initial results are very promising and open up a way to formally mathematically define the concept of structural equivalence. However, further deeper study on this matter must be carefully conducted in order to confirm the viability of our proposed method. For instance, if we relax the condition to approximately equal then nodes 80 and 83 from Table 7 will also be considered equivalent.

Finally, we have also tested and confirmed the hypothesis of Granovetter. Consider two arbitrary selected individuals A and B and the set of all persons with ties to either or both of them. The hypothesis is: the stronger the tie between A and B, the larger the proportion of individuals in S to whom they will be both tied (connected by a weak or strong tie).  For Group 1, there are a total of 5526 strong wedges out of which: 39.052 \% are closed by another strong link, 33.406 \% are closed by a weak link and 72.457 \% are closed by any link. Also, there are a total of 29860 weak wedges out of which: 8.279 \% are closed by a strong link, 9.752 \% are closed by another weak link and 18.031 \% are closed by any link. For Group 2, there are a total of 4811 strong wedges out of which: 39.243 \% are closed by another strong link, 45.022 \% are closed by a weak link and 84.265 \% are closed by any link. Also, there are a total of 25424 weak wedges out of which: 14.301 \% are closed by a strong link, 23.450 \% are closed by another weak link and 37.752 \% are closed by any link.

\begin{table}[!ht]
\caption{{\bf Structural equivalence. Graph characteristics of the nodes in Group 1 with $d_i^{out} = 4$ and $d_i^{in} = 4$ }}
\begin{tabular}{|l|l|l|l|}
\hline
{\bf strong OFF}&  $r_i$  & $tc_i$ &  $tp_i$ \\ \hline
$i=2$ & 1 & 0.424 &	0.442  \\ \hline 
$i=40$	& 0.333	& 0.042 & 0.094	 \\ \hline 
$i=80$ & 0.6 & 0.343 & 0.353	 \\ \hline 
$i=83$ & 0.6 & 0.356 & 0.287	 \\ \hline 
$i=94$	& 0.6 & 0.053 & 0.056	 \\ \hline 
$i=122$ & 1 & 0.376 & 0.366	 \\ \hline 
$i=127$ & 1 & 0.424 & 0.442	 \\ \hline   
$i=148$ & 0.143 & 0.083 & 0.191	 \\ \hline 
\end{tabular}
\end{table}

\section{Conclusions}

By studying friendship relations among students enrolled on two different courses represented using a multiplex structure, a number of interesting conclusions regarding the strength of online and offline ties can be drawn: (1) strong ties are preferred in face-to-face (offline) communications; (2) weak ties are equally presented in online and offline communications; (3) in offline communication, strong and weak ties are (almost) equally included; (4) in online communication weak ties are dominant; (5) weak ties (in three layers of the multiplex network: weak offline, weak online, and weak) are much less reciprocal than strong ties; (6) dyads (that is, reciprocities) are preserved, however triads (measured with normalized three cycles and triplets) are not significant in different layers of a multiplex network; (7) activity and popularity patterns for some layers of the multiplex network are significant: out-degree and in-degree friends in one layer could also be out-degree and in-degree friends in another layer. These conclusions are supported by the data obtained from both groups. The number and consistency of the drawn conclusions have confirmed that by approaching the problem of different interrelationships between actors as a multiplex network problem, one can gain useful insight on the importance of each type of link within the social network as a whole, as well as on the interaction and overlapping between different link types, especially in the offline/online (real/virtual) environments as was our case of study.

In the future, we plan to study how exogenous characteristics influence the structure of the multiplex network and how the network structure dominates the actors (students) characteristics and their temporal evolution. In particular, we will address questions on: how student grades are distributed; students' commitment and progress to the studies; how one could empower the students to obtain better grades or determine their specific areas of interest with greater success. Are the students with better grades and habits more central to the network or not, and how changing the placement of these individuals could influence the structure of the network?

%
%
%

\end{document}